\documentclass{PoS}

\usepackage[sort&compress,numbers]{natbib}
\usepackage{amsmath}
\newcommand{\bo}{\ensuremath{\boldmath{B}_0}}
\newcommand{\ez}{\ensuremath{{\boldmath{\hat e}}_z}}
\newcommand{\f}[1]{\ensuremath{\boldmath{#1}}}
\newcommand{\pa}{\ensuremath{_\parallel}}

\newcommand{\df}{\ensuremath{\mathrm{d}}}
\newcommand{\Om}{\ensuremath{\varOmega}}
\newcommand{\bi}{\begin{itemize}}
\newcommand{\ei}{\end{itemize}}

\title{Cosmic-ray diffusion in magnetized turbulence}

\ShortTitle{Cosmic-ray diffusion}

\author{\speaker{Robert C. Tautz}\\
Zentrum f\"ur Astronomie und Astrophysik, Technische Universit\"at Berlin, Hardenbergstra\ss{}e 36, D--10623 Berlin, Germany\\
E-mail: \email{robert.c.tautz@gmail.com}}

\abstract{The problem of cosmic-ray scattering in the turbulent electromagnetic fields of the interstellar medium and the solar wind is of great importance due to the variety of applications of the resulting diffusion coefficients. Examples are diffusive shock acceleration, cosmic-ray observations, and, in the solar system, the propagation of coronal mass ejections. In recent years, it was found that the simple diffusive motion that had been assumed for decades is often in disagreement both with numerical and observational results. Here, an overview is given of the interaction processes of cosmic rays and turbulent electromagnetic fields. First, the formation of turbulent fields due to plasma instabilities is treated, where especially the non-linear behavior of the resulting unstable wave modes is discussed. Second, the analytical and the numerical side of high-energy particle propagation will be reviewed by presenting non-linear analytical theories and Monte-Carlo simulations. For the example of the solar 
wind, the impact of anisotropic and dynamical turbulence models will be discussed. In addition, it will be shown how further complications can be treated that arise from the large-scale magnetic field geometry and turbulent electric fields. The transport properties of energetic particles can thus be calculated for current turbulence models so that they withstand a comparison with measurements taken in the solar wind.}

\FullConference{Cosmic Rays and the InterStellar Medium - CRISM 2014,\\
24-27 June 2014\\
Montpellier, France}

\begin{document}

\section{Introduction}

Cosmic-ray research offers a plethora of questions ranging from basic physical principles to aspects that are immediately related to man and technology. Both the understanding and, more tangibly, the prediction and explanation of measured data concerning the so-called ``space weather'' \citep{sch05:wea} therefore cannot be overestimated. Owing to a continuing technological development of near-Earth space, such research is also connected to economic interests.


While the basic history of cosmic rays and their discovery is part of introductory courses in astronomy, there are many interesting, less well-known facts that are now summarized:
(i) the discovery of cosmic rays could in fact have been made before, as there were some people who noticed radiation even in well-shielded ionization chambers or who undertook balloon flights but failed to interpret the increasing radiation as being of ``cosmic'' origin;
(ii) apart from known periodicities such as the solar modulation, additional variations in the cosmic-ray intensity seem to be very rare;
(iii) A first, both systematic and successful, investigation was made by Viktor Hess, who concluded that, in addition to terrestrial and atmospheric radiation, there must be a third, cosmic, source in order to explain his measurements;
(iv) had Viktor Hess been able to climb to a considerably higher altitude, he would have been in trouble explaining the now decreasing ionization rate. Today we know that, in fact, he had measured secondary particles created in the upper atmosphere instead of the primary, truly ``cosmic'' particles;
(v) the term ``cosmic rays'' was coined by Gockel and Wulf and later, independently, by Millikan \citep{wal12:hoh};
(vi) Fermi was the first to consider theoretically the acceleration processes of cosmic-ray particles and to propose their extra-solar origin. Nowadays, the same principle is also used to heat laboratory plasmas.

Independently, the history of astrophysical magnetic fields started in 1908, when George E.{} Hale successfully measured the Zeeman splitting of solar absorption lines and also managed to estimate the field strength. Galactic magnetic fields have been investigated for the past 45~years. Measurements are considerably more difficult than for the Earth or solar magnetic fields and proceed as follows:
(i) even though typical galactic magnetic field strengths are of the order of only a few micro-Gauss, the Zeeman effect can still be discerned from Doppler shifting and thermal broadening for example in \textsc{Hi}~regions;
(ii) synchrotron emission by electrons spiraling around magnetic field lines. However, the electron velocity must be relativistic, thereby allowing for a sufficiently bright intensity so that the radiation can be measured with Earth-bound radio telescopes;
(iii) polarization measurements indicate macroscopic dust particles that align themselves in a magnetic field so that their thermal emission becomes polarized;
(iv) Faraday rotation measurements exploit a peculiar plasma effect, due to which right-hand and left-hand polarized waves propagate with different group velocities. Because the effect is sensitive to the product of electron density and magnetic field strength, the latter can be estimated as soon as the density is known from other measurements such as the dispersion measure.

In general, galactic magnetic fields can be arranged in two groups, which are (i) large-scale, or regular, magnetic fields which are varying on scales larger than 1000~parsecs and produce polarized radio emission \citep{bec96:mag}; (ii) turbulent magnetic fields with scale lengths below 100~parsecs, parts of which may have been originated in plasma instabilities that arise in the interstellar medium.


For a long time, the fields of plasma physics and that of astrophysics have developed separately due to historical reasons. Therefore, many researchers have specialized in either astrophysics---mostly with applications to stellar evolution and the dynamics of galaxies, to mention only two examples---or plasma physics, with applications to fusion devices and other laboratory plasmas. Only recently have there been attempts to reproduce astrophysical turbulence in the laboratory. 


For plasmas, a theoretical description based on (magneto) hydrodynamics is often insufficient because of long-range electromagnetic forces, thus requiring a high particle density for the MHD description to be valid. More precisely, electromagnetic interactions are \emph{not} allowed to dominate Coulomb collisions because such collisions are responsible for the formation of a Maxwellian velocity distribution, on which the equations for pressure and temperature are based. Hence the requirement of frequent particle collisions. In contrast, for more tenuous plasmas the particle distribution in velocity space is generally unknown, and a more general approach has to be chosen. Such is represented by the kinetic plasma equations, which combine Maxwell's equations with the continuity equation in phase space, better known as the Vlasov or collisionless Boltzmann equation.


Solving the Vlasov equation together with Maxwell's equations is a major challenge, because both are coupled through the charge and current densities, which are given by the zeroth and first moment of the distribution function, respectively. Therefore, three basic approaches have mainly been used:

\bi
\item The \emph{test-wave approach} is used to obtain the plasma turbulence generated by the particles, while at the same time the influence of the turbulent fields on the distribution function is neglected. The Maxwell-Vlasov equations are linearized around a given initial equilibrium such as a (possibly anisotropic) Maxwellian together with a homogeneous magnetic field. Together with the generalized Ohm's law, a subsequent Fourier-Laplace transformation introduces the wave vector and the frequency. The inverse Laplace transformation finally allows one to find dispersion relations for the plasma waves by exploiting the residue theorem in the limit of large times.

\item The \emph{test-particle approach} is used to investigate how charged particles behave in a given electromagnetic turbulence. To do so, the ensemble-averaged distribution function is used to calculate the fluctuating part. Such allows one to transform the Vlasov-equation to a Fokker-Planck equation, which, in application to energetic particle transport in the interplanetary medium, reduces to the Parker transport equation. Similarly, in a second step, the isotropic part of the distribution is used to express the anisotropic part, thereby yielding a diffusion-convection equation (for an introduction, see \citep{rs:rays,sha09:nli} and references therein).

\item In \emph{numerical simulations}, the system of the Vlasov-Maxwell equations can, in principle, be solved self-consistently. However, it has become convenient to prescribe an initial state that corresponds either to the test-wave approach (to verify a stability analysis) or to the test-particle approach (to judge the analytical determination of scattering parameters).
\ei

\section{Plasma instabilities}

Historically, the term ``instability'' results from early research on nuclear fusion, where one was trying to keep a plasma magnetically confined. Here, instabilities were unwanted, because they tended to destroy the confinement. One the one hand, the management of instabilities in fusion devices continues to be important. In astrophysics, on the other hand, plasma instabilities are a fascinating phenomenon, because they are able to create electromagnetic fields from scratch.

An important class of instabilities operates in initially unmagnetized plasmas, where the background magnetic field is zero. If, additionally, an instability has a vanishing wave frequency and has only a positive growth rates, it is called ``aperiodic'', because it does not propagate but simply grows in place. Such instabilities are generally referred to as a ``Weibel-type'' instability in honor of their discoverer Erich S.{} Weibel. The original derivation of the Weibel instability, however, operates specifically in a plasma at rest with only a temperature anisotropy \citep{wei59:wei}.


\subsection{Role of the Distribution Function}\label{ins:dist}

Most investigations discussed in the literature focus on specific forms of the particle momentum (or velocity) distribution. Examples are: (i) Weibel's original assumption of an anisotropic Maxwellian distribution for a warm plasma at rest; (ii) cold counterstreaming plasma components without thermal broadening; (iii) spatial temperature gradients. Combinations are, of course, also possible and are frequently used for stability analyses \citep{tau11:max}.

A non-exhaustive list of more specialized distribution functions that have been used over the years includes:
(i) the so-called waterbag distribution, which is used to provide the simplest possible form of thermal spread without a functional dependence on the velocity variables to the integrals;
(ii) Cauchy-type distributions, which resemble the Gaussian form but, at the same time, are easier to be dealt with analytically due to their fractional polynomial structure;
(iii) kappa-type distributions, which have been expected to be found in dilute plasmas and have been measured in interplanetary plasmas. Consequently, such distributions were subject to many instability analyses;
(iv) so-called ``semi-relativistic'' distributions that allow for relativistic temperatures in one direction, while for the other directions non-relativistic temperatures are assumed;
(v) a fully relativistic, two-dimensional counterstreaming distribution in velocity space as required, for instance, by particle-in-cell codes;
(vi) anisotropic, fully relativistic generalizations of the Maxwell-Boltzmann-J\"uttner distribution in momentum space, which is evidently the generalization of the Maxwellian distribution for thermal velocities close to the speed of light.

An overview of the modifications to the Weibel instability due to changes in the distribution function has been given by \citet{usr08:wei}.


\subsection{Non-linear extensions}

Weibel modes and their associated non-linear structures play a role for many physical effects. The relevant extensions of the Weibel instability, other instabilities including the Harris instability, and both their respective linear and non-linear aspects include (for an overview, see \citep{tau12:rad} and references therein):
(i) relativistic factors (both with and without external magnetic field);
(ii) coupling factors of the transverse and longitudinal wave behaviors;
(iii) generation of magnetic fields from the currents;
(iv) modification of Weibel-type instabilities due to a background magnetic field;
(v) specialized investigations of beam-plasma interactions and counterstreaming plasmas;
(vi) saturation of the Weibel instability by non-linear effects;
(vii) particle-in-cell calculations to investigate Weibel-like instabilities;
(viii) Weibel-like behavior in quantum plasmas;
(ix) the combination of unstable plasma distributions and particle scattering properties;
(x) and a host of ancillary factors including multiple species distribution functions that may be partially anisotropic per species or isotropic (but not for all species).

A further aspect of the Weibel instability and its later extensions is the influence of an ambient magnetic field on the instability both in terms of the linear wave regime and the non-linear soliton regime. The presence of an ambient magnetic field tends to make particle distributions gyrotropic in the plane perpendicular to the field direction because of the circular motion around the field. However, such is by no means the same as having the particle distribution gyrotropic relative to the wave propagation direction, unless the wave runs parallel to the field.


\subsection{Magnetized turbulence}

The classic turbulence theory \citep{bat82:tur}, which was formulated for velocity disturbances in a fluid or a gas, can in principle be adapted to describe turbulent electromagnetic fields. In addition to the twisting of large-scale magnetic fields, such turbulent fields are thought to be generated by plasma instabilities.
%
%
In order to allow for a simplified description of the turbulence geometry, typically incompressible homogeneous turbulence is assumed, although there have been attempts to include the effects of intermittency and other large-scale variations.
%
%
Four prominent models have been used both in analytical and numerical investigations, and even attempts at more realistic models are based on these basic layouts, which are:

\bi
\item In the \emph{slab model}, the wave vectors of the turbulent magnetic fields are always aligned with the background magnetic field, \bo. Assuming that $\bo=B_0\ez$, one has, therefore, $\delta\f B=\delta\f B(z)$. In most cases, slab turbulence leads to sub-diffusive perpendicular diffusion coefficients with $\kappa(t)\propto t^{-1/2}$, where ``perpendicular'' refers to the $x$-$y$ plane. Perpendicular diffusion is also recovered if wave propagation effects are taken into account. Based on Mariner~2 measurements it was found that, to some extent, the solar wind is dominated by an outward propagating Alfv\'enic turbulence, which is believed to be the origin of the slab component;

\item The opposite case is that of \emph{2D turbulence}, where $\delta\f B=\delta\f B(x,y)$. For this case, theories have been developed that relate the pitch-angle Fokker-Planck coefficient to the perpendicular diffusion coefficient. By assuming diffusive cross-field transport, sub-diffusive pitch-angle scattering was found that lead to super-diffusive particle transport along the mean field, i.\,e., $\kappa\pa(t)\to\infty$. This second turbulence contribution may originate from aperiodic fluctuations, which assumption is supported by the observation that the two-dimensional component appears to be time-independent;

\item The \emph{composite} or \emph{two-component turbulence} model is the combination of slab and 2D turbulence, where originally 80\% 2D is completed by 20\% slab turbulence \citep{bie94:pal}. Later, newer methods to derive this ratio from magnetometer data of the \textsc{Helios}~1 and 2 spacecraft taken at radial distances from $0.3$ to 1\,a.u.{} changed the ratio to 85\% 2D and 15\% slab;

\item In the \emph{isotropic model}, it is assumed that no preferred direction exists for the turbulent magnetic field. In this case the parallel diffusion coefficient (if transport along the mean field is indeed diffusive) is controlled by the pitch-angle diffusion coefficient close to $90^\circ$. Therefore, quasi-linear theory, which cannot describe $90^\circ$ scattering correctly, has to be replaced by a non-linear diffusion theory. Thus a second-order theory was used to compute parallel diffusion coefficients in isotropic turbulence that are in agreement with numerical simulations \citep{tau08:soq}. In general, a theoretical description of cross-field transport in isotropic turbulence is difficult.
\ei

There are more advanced anisotropic models, most of which are based on sophisticated numerical and analytical investigations of turbulence. Strictly speaking, the Kolmogorov ansatz is only partially valid for MHD turbulence so that the spectral index may be anisotropic. A famous example is the Goldreich-Sridhar turbulence model \citet{sri94:tur}, which takes into account the modification of the energy spectrum due to Alfv\'en waves. This naturally gives rise to anisotropy in the directions parallel and perpendicular to the mean magnetic field.

\section{Test-particle transport}

The fact that cosmic rays appear to be distributed isotropically over the entire sky is important because, at first glance, it is deceiving. Understanding the measurements of cosmic rays made on Earth requires a precise understanding of their motion through space. Because cosmic ray particles are charged, they are subject to deflection by all kinds of electromagnetic fields found everywhere in extragalactic space, in the interstellar medium, and in the solar system. The description of such particle motions is a subject of wide interest, because is important applications also in laboratory plasmas. Nevertheless, it is afflicted with many difficulties.


\subsection{Analytical transport theory}

While the basic question is simple---determine the particle motion in turbulent electromagnetic fields without taking into account the back-reaction of the particles on the fields---no conclusive answer has been found so far except for very simplified turbulence models. Usually, a diffusive behavior of the particle motion is assumed so that the formalism of diffusion can be applied.


However, it has been demonstrated that the standard theory of cosmic-ray diffusion, the ``classic'' quasi-linear theory (QLT) \citep{jok66:qlt}, often yields invalid predictions or results in singularities for time-independent magnetic turbulence. The reason is that it cannot describe the so-called 90$^\circ$~scattering, where particles reverse their motion in the direction parallel to the ambient magnetic field \citep{tau06:sta}. To solve this problem, a number of non-linear theories have been proposed, some of which actually give an accurate description of the transport parameters.

A few selected, contemporary non-linear theories (for an overview, see \citep{tau12:nov} and references therein) are:

\bi
\item The \emph{compound-diffusion model}, which first determines the mean square displacement of magnetic field lines, for which quasi-linear theory turned out to be surprisingly accurate. The particle diffusion coefficient is then obtained from a convolution integral of the field-line diffusion coefficient together with a suitable distribution function such as the solution of the diffusion equation. The advantage is that only the second step is ``non-linear'', which simplifies the underlying calculations. Hence the name \emph{``semi--quasi-linear theory''};

\item The \emph{second-order quasi-linear theory (SOQLT)}, in which the key assumption of quasi-linear theory---undisturbed particle trajectories and therefore resonance with a countable number of discrete wave modes---is replaced by resonance broadening. Whereas the analytical evaluation of the equations for fully three-dimensional turbulence is difficult, simplified expressions can be obtained for the case of slab turbulence. Furthermore, the application to interstellar cosmic-ray transport indicates that particles with Larmor radii larger than the diameter of the Galaxy---which is known as the Hillas limit---particles might still be confined to the Galaxy \citep{sha09:hil};

\item The \emph{non-linear guiding center (NLGC)} theory of \citet{mat03:nlg}, which is based on the magnetic field-line equation. In the original formulation, some questionable mathematical approximations had been used, but the results nevertheless were in good agreement with numerical simulations. Since then, several non-linear extensions  have been suggested, which improved both the theoretical foundations as well as the accuracy;

\item The \emph{unified non-linear theory (UNLT)}, which explicitly calculates the ensemble average by employing the solution of the Fokker-Planck equation as the weighting function. For the first time, the theory is able to provide the correct sub-diffusive perpendicular mean-free path in slab turbulence \citep{sha11:sub}. There remain, however, several aspects in the basic mathematical formulation that require additional clarification.
\ei

In addition, a number of heuristic approaches such as the BAM model (see the references in \citep{sha09:nli}) or the interpolation formulae of \citet{rei93:mod} try to condense the complex physics of a quasi-diffusive motion to as simple as possible an expression. Such has the advantage that further mathematical steps remain analytically tractable.


\subsection{Additional effects}

Most basic transport theories assume: (i) a homogeneous background magnetic field, (ii) an axisymmetric turbulence geometry; (iii) a simplified turbulence description which in most cases neglects a ``slow'' time-dependence such as large-scale variations in the turbulence sources or wave-damping effects. Examples for the inclusion of additional features are:
(i) the off-axis elements of the diffusion tensor, which can be shown to correspond to the classic drift motions that occur in a plasma whenever the magnetic field is curved;
(ii) a curved background magnetic field as defined via a so-called adiabatic ``focusing length'' via $L=-B/(\df B/\df s)$. The derivative is to be taken along the magnetic field \citep{roe69:int}. However, such enforces a coherent particle motion towards weaker magnetic fields, thereby obscuring the diffusive motion \citep{tau14:adf};
(iii) dynamical effects \citep{bie94:pal} and intermittency as well as the explicit incorporation of various types of plasma waves. Such allows one to work out the effects of momentum diffusion and stochastic acceleration;
(iv) the influence of anisotropic plasma-wave polarization or non-vanishing magnetic helicity on the random walk of magnetic field lines, which has been investigated both analytically and using test-particle simulations.

A fundamental problem, however, is the fact that each such effect can have a dramatic influence on the resulting transport parameters. An example is the diffusivity of the particle motion \citep{tau10:sub}. While the parallel (perpendicular) mean free paths are diffusive (sub-diffusive) for magnetostatic slab turbulence, they become super-diffusive (diffusive) in the case of plasma-wave turbulence with electric fields included. A slowly varying mean magnetic field as for example in the Parker spiral may also lead to the recovery of perpendicular diffusion even in the case of magnetostatic turbulence. Predictions as to which effects might be neglected are therefore often untrustworthy.


\subsection{Numerical simulations}

In \emph{Monte-Carlo \emph{or} test-particle simulations} (see \citep{tau13:num} and references therein), the large-scale magnetic fields are prescribed at the initial stage like for particle-in-cell simulations. But in addition, the turbulent fields are now also prescribed through their statistical properties to ensure the comparability to analytical test-particle theories. The underlying parameters that have to be specified are: (i) the power spectrum of the turbulence; (ii) the time dependence (magnetostatic turbulence, plasma waves, dynamically decaying turbulence); and (iii) the geometry of the turbulent magnetic fields. This approach allows for the investigation of transport parameters such as the mean free path and the diffusion coefficients. The advantage in comparison to test-particle calculations is that now no uncertainties are involved that could arise from the use of analytical approximations.


To ensure proper scattering of the test particles, two basic conditions have to be fulfilled:
(i) the \emph{resonance condition} states that there has to be one wavenumber $k$ so that $R_{\text L}k\approx1$, where $R_{\text L}$ denotes the particle's Larmor radius. Scattering, therefore, predominantly occurs when a particle can interact with a wave mode over a full gyration cycle;
(ii) the \emph{scaling condition} requires that $R_{\text L}\Om_{\text{rel}}t<L_{\text{max}}$, where $L_{\text{max}}\propto1/k_{\text{min}}$ is the maximum size of the system, which is given by the lowest wavenumber (for which one has $k_{\text{min}}=2\pi/\lambda_{\text{max}}$, thereby proving the argument). In practice, this condition determines the minimum wavenumber the turbulence generation begins with.

Only if both conditions are fulfilled will the particles be properly scattered. If, for example, the scaling condition was violated, the particles would appear to be free-streaming as if there were no turbulent fields. Such can be understood by remembering that, although the turbulence is still generated at every particle position, the system behaves as if it had a finite box size given by $L_{\text{max}}$. Once the particle leaves that box, it is no longer part of the turbulent system.

\section{Solar system}

One of the most prominent astrophysical scenarios with both turbulent and curved large-scale background magnetic fields is the solar system \citep{bal13:hel}. Here, the ambient field is represented by the Sun's magnetic field, which is shaped in the form of the famous Parker spiral, superposed by a turbulent component generated mainly by the solar wind. The main advantage is that, inside the solar system, in-situ measurements can be done of both electromagnetic fields and energetic particles \citep{bak12:spa}, even in coincidence.


Solar particles populate the relatively low energy-range of the cosmic-ray energy spectrum. Here, the influence of the dissipation range in the turbulence power spectrum becomes increasingly important and must, therefore, be accounted for in the analytical model. Conveniently, measurements by Cluster are available \citep{hor08:mul}, which indicate a value of approximately $p\approx2.9$ for the dissipation range spectral index (note that the symbol~$p$ is customary for the dissipation range spectral index, even though it also denotes the momentum).

A recent careful investigation shows that more realistic, fully three-dimensional models such as the Maltese cross \citep{mat90:mal,rau13:mal} are required because the resulting diffusion parameters significantly differ from that obtained with the composite model. The inclusion of plasma waves for the slab contribution in the composite model results in excellent agreement \citep{tau13:pal} with observational data.

Apart from such large-scale, fully developed turbulent fields, there are also specific examples where the generation of turbulent fields via plasma instabilities can be observed. The collected WIND/SWE data falls well within the stability limits of several known plasma instabilities. Our reasonable understanding of astrophysical plasma theory is thus underscored, although some aspects are still unclear \citep{sch11:thr}.

\section{Conclusion}

The current theoretical framework for the description of turbulence and wave-particle interaction in astrophysical plasmas has enabled many great achievements. Two basic tools of kinetic theory are the test-wave approach and the test-particle approach, which describe two complementary ways of investigating the same basic problem, namely to solve the coupled system of Maxwell's equations together with the Vlasov equation. For their respective original areas of application, both have been verified many times and have helped to understand a variety of astrophysical phenomena that otherwise would have remained inscrutable.

Nevertheless, significant open questions remain to be answered. These include: How can non-linear plasma instabilities be combined with particle transport theories?  How do scattering theories need to be revised according to new results in turbulence theory and observations? How does small-scale cosmic-ray induced turbulence connect to large-scale hydrodynamic or MHD turbulence? How do cosmic rays influence the evolution of the galaxy as a whole?

A major challenge for future work is the combination of the individual approaches in order to allow for the investigation of phenomena that require a fully self-consistent treatment. In my view, efforts should be focused both on the proper understanding and application of standard physics, which sometimes might not be as spectacular as the invention of ``new'' physics. However, modern observations provide a huge amount of fascinating data, which needs to be interpreted and will surely continue to overthrow many currently accepted theories.

\acknowledgments

I thank Alexandre Marcowith for the invitation to the CRISM 2014 conference. In addition, I thank everyone who supported me during my \emph{Habilitation} two years ago.

\small

\end{document}